\documentclass[12pt,preprint]{aastex}

\usepackage{mathptmx}
\usepackage{courier}
\usepackage{graphicx}
\usepackage{colordvi}
\usepackage{natbib}
\usepackage{color}
\usepackage{wasysym}

\shorttitle{Brightness correction to SoHO MDI continuum images}
\shortauthors{Criscuoli et al.}

\begin{document}

   \title {Line shape effects on intensity measurements of solar features:  \\
Brightness correction to SoHO
    MDI continuum images
    }
   \author{S.~Criscuoli\altaffilmark{1},  I.~Ermolli\altaffilmark{1}, D.~Del~Moro\altaffilmark{2}, F.~Giorgi\altaffilmark{1},  A.~Tritschler\altaffilmark{3}, H.~Uitenbroek\altaffilmark{3} and N.~Vitas\altaffilmark{4} }

    \altaffiltext{1}{INAF-Osservatorio astronomico di Roma, Via Frascati 33, 00040 Monte Porzio Catone, Italy; serena.criscuoli@oaroma.inaf.it}
\altaffiltext{2}{ Department of Physics, University of Roma Tor Vergata, Via della Ricerca Scientifica 1, I-00133 Roma, Italy}
\altaffiltext{3}{ National Solar Observatory, Sacramento Peak, P.O. Box 62, Sunpsot, NM 88349, USA}
\altaffiltext{4}{Sterrekundig Instituut, Utrecht University, Postbus 80 000, 3508 TA Utrecht, The Netherlands}

   \date{}

  \begin{abstract}
Continuum intensity observations obtained with the Michelson Doppler Imager (MDI) on-board the SoHO mission  
provide long time series of filtergrams that are ideal for
studying the evolution of large-scale phenomena in the solar atmosphere and their dependence on solar activity.
These filtergrams, however, are not taken in a pure continuum spectral band, but are constructed from a
proxy, namely a combination of filtergrams sampling the \ion{Ni}{1} 6768 \AA\  line.
We studied the sensitivity of this continuum proxy to the shape of the nickel line and to the degradation in the instrumental
transmission profiles. 
We compared continuum intensity measurements in the nearby of nickel line with MDI proxy values in three sets of
high resolution spectro-polarimetric data obtained with the Interferometric Bidimensional Spectrometer
(IBIS), and in synthetic data, obtained from
multi-dimensional simulations of magneto-convection and one-dimensional atmosphere models.
We found that MDI continuum measurements require brightness corrections which depend on magnetic
field strength, temperature and, to a smaller extent, plasma velocity.
The correction ranges from 2\% to 25\% in sunspots, and is, on average, less than 2\% for other features.
The brightness correction also varies with position on the disk, with larger variations obtained
for sunspots, and smaller variations obtained for quiet sun, faculae and micropores. 
Correction factors derived from observations agree with those deduced from the numerical simulations
when observational effects are taken into account.
Finally, we found that the investigated potential uncertainties in the transmission characteristics of MDI filters 
only slightly affect the brightness correction to proxy measurements.

\end{abstract}


\maketitle

\section{Introduction}
The continuum intensity measurements obtained with MDI
have provided extensive time series of data 
unaffected by seeing for over 13 years.
These data are ideal for studying large-scale phenomena in the solar atmosphere 
and their dependence on solar magnetic activity. 
They have been utilized in many investigations, concerning the analysis of 
large-scale patterns in plasma motions 
 \citep[e.g.,][and references therein]{meunier2008,meunier2007}, 
the measurement of the radiative properties of magnetic 
elements  over the activity cycle \citep[e.g.][]{ortiz2002,ortiz2006,mathew2007},  
and the modeling of irradiance variations \citep[e.g.][]{krivova2003,wenzler2006,wenzler2009}.

Nevertheless, MDI continuum data are a by-product of the
instrument, which was designed mainly for Doppler
measurements. Specifically, intensities in the continuum are derived
by the combination of five narrow-band filtergrams, obtained with filters sampling a passband of 94 m\AA\ FWHM, equally spaced by 75 m\AA\ around the \ion{Ni}{1} 6768 \AA\ mid-photospheric line.  
The filtergrams are labeled $F_0 \ldots F_4$, where $F_0$
is divided over two bands taken near the continuum, on either side of the line,
$F_1$ and $F_4$ are centered on the wings of the nickel line, and
$F_2$ and $F_3$ are centered around its core. 
Even at $F_0$ the instrument does not sample true continuum.
Instead, a proxy continuum-intensity filtergram is
constructed by combining the five nominal filtergrams in the following way
\citep{scherrer1995}:
\begin{equation}
  I_c = 2 F_0 + I_{\rm{depth}}/2 + I_{\rm{ave}},
  \label{eq:IcontMDI}
\end{equation}
where $I_{\rm{ave}}$ is the average of the five filtergrams and $I_{\rm{depth}}$ is the line depth,
which is given by:
\begin{equation}
  I_{\rm{depth}} = \sqrt{2((F_1 - F_3)^2 + (F_2 - F_4)^2)}.
  \label{eq:linedepth}
\end{equation}
The components of sum $I_c$ theoretically have cancelling systematic errors as a 
function of solar velocity, so that the continuum image is claimed to be free of
Doppler induced cross-talk at the 0.2\% level \citep{scherrer1995}.

The accuracy of the MDI continuum estimate (called the MDI method hereafter) is, however,
inherently limited by the shape of the spectral line and its
sensitivity to both strength and inclination of the magnetic field, local
thermal stratification, and line-of-sight motions of the observed region.  
In addition, MDI measurements may also suffer from substantial errors
because of uncertainties in the actual transmission characteristics of the filters utilized
\citep{wachter2008}. 

Several previous studies have focused on the accuracy of magnetic flux estimates obtained from MDI observations
\citep[e.g.][]{berger2003,tran2005,ulrich2009,wang2009, demidov2009,zhendong2010},
as well as the accuracy of dynamic measurements \citep[][]{wachter2006,rajaguru2007,wachter2008}.
Results indicate a substantial underestimate of magnetic flux and
spurious contributions affecting Doppler measurements.
This lead to recalibration of magnetograms and dopplergrams series, with the most  
recent results suggesting that the calibration of these data could be improved yet further.
By contrast, the accuracy of continuum intensity measurements has only been investigated by
\citet{mathew2007}, using spectral calculations based on one-dimensional atmospheric models. 
Their results indicate that MDI continuum measurements underestimate intensity in 
sunspots, with an error depending on both the magnetic field strength and temperature stratification. 

The objective of this study is to further investigate the uncertainties affecting MDI continuum intensity measurements,
by considering both high spatial resolution spectro-polarimetric observations and numerical simulations.
To this aim we defined a brightness correction factor and investigated its variation with the physical properties of the
analyzed solar features and observational conditions, such as spatial and spectral scattered light,
finite spatial and spectral resolution and position on the solar disk. We also investigated possible effects 
resulting from degradation of MDI filter transmission profiles.

The paper is organized as follows. We describe the data analyzed and provide some details of the 
reduction applied (Sect. 2). Then,  
we present  
the spectral synthesis method and the results obtained from our measurements and synthesis computations (Sect. 3). We also investigate sensitivity of results to filter profiles uncertainties (Sect. 4). We finally discuss 
the results obtained and present our conclusions (Sect. 5).


\section{Spectral data}
To investigate the sensitivity of the MDI continuum measurement proxy we employed profiles of the 
\ion{Ni}{1} 6768 \AA\ line obtained from two different sources, namely high-resolution spectro-polarimetric
observations acquired with the Interferometric Bidimensional Spectrometer \citep[IBIS][]{cavallini2006,reardon2008}
at the NSO/Dunn Solar Telescope, as well as with theoretical line synthesis calculations, as described below.

\begin{figure}[!ht] 
\includegraphics[width=9cm]{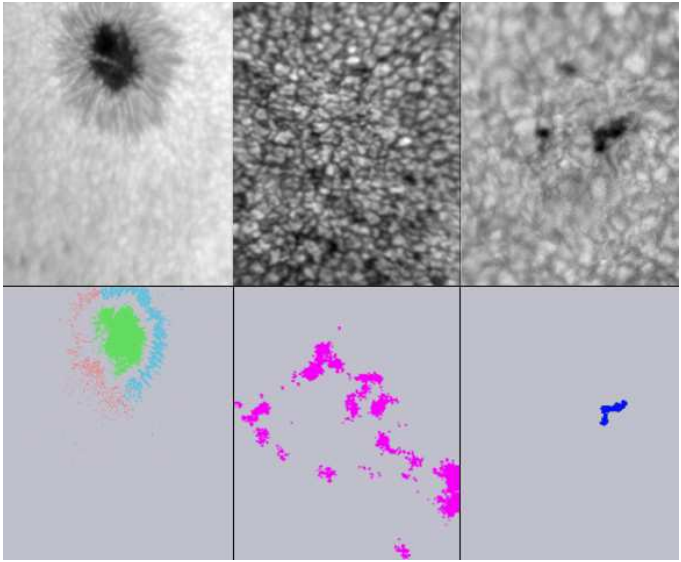}
\caption{ Top: subfields of 40 arcsec $\times$ 50 arcsec of frames from \textit{Sunspot} (left), \textit{Faint Plage}
 (middle) and \textit{Pore} (right) data sets. Bottom: masks depicting features identified according to criteria in 
Table~\ref{tabellaid}. The colour code is the same as in Fig. \ref {fig_prof_obs}.
} 
\label{fig1} 
\end{figure}   

\subsection{Observed spectra}
The observed spectra used in this work were extracted from three
IBIS data sets.  These will be referred to as the \textit{Pore},
\textit{Sunspot}, and \textit{Faint Plage} sets hereafter, alluding
to the main solar features in their field-of-view (FOV).  The \textit{Pore} set
is a spectral scan of the \ion{Ni}{1} 6768 \AA\ line acquired near
the disk centre with diffraction limited spatial sampling at
0\farcs083/pixel.  The \textit{Sunspot} and the \textit{Faint Plage}
sets are Full-Stokes spectropolarimetric scan of the \ion{Ni}{1}
6768 line, acquired at $\mu = 0.88$ and $\mu = 0.92$, respectively,
with a spatial sampling of 0\farcs17/pixel.  The IBIS spectral profile
at 6768 \AA\ has a FWHM of 25 m\AA\ \citep{reardon2008} and the
spectral sampling was about 30 m\AA\ for the three sets.
The datasets were reduced following standard procedures
\citep{cauzzi2008, viticchie2009, judge2010} which also correct for
the instrumental blue-shift, and for instrument and telescope
polarization in the case of spectropolarimetric data.  The pipeline
includes a MFBD image reconstruction technique \citep{vannoort2006}
to reduce seeing degradation and possible seeing induced
crosstalk.

Note that all observed line profiles suffer from 
finite spectral resolution and spectral scattered light contamination.
This last effect was estimated by comparison of observed average quiet Sun
profile with Atlas measurements by \citet{kurucz1984}, following the procedure
described in \citet[][their eq.~1] {cabsol2007}.
We found a consistent light contamination level of 12\% on all the three datasets analyzed.

To estimate the magnetic flux density along the Line-of-Sight ($B_{LOS}$) 
in image pixels of the \textit{Sunspot} and \textit{Faint Plage} sets, we applied the Center Of Gravity method (COG) by
\citet{rees1979}. This method provides a reasonable estimate of the $B_{LOS}$
at the formation height of the line core as demonstrated by \citet{uitenbroek2003}
for several lines of \ion{Fe}{1}. 
We applied the method to synthetic sunspot profiles from the 3D MHD simulations (described in Sect. 2.2) taking into account both spatial and spectral scattered light
and found that the COG in the \ion{Ni}{1} line
can underestimate magnetic flux by up  to $\approx$ 28\% for $B_{LOS} > $1.5 kG. 
The $B_{LOS}$ value deduced for each image pixel was compensated for LOS projection by applying the relation
\citep[][p.\ 660]{Landi+Landolfi2004}
\begin{equation}
B = B_{LOS}/\cos{\gamma}  
\label{formula1}
\end {equation}
where the inclination angle $\gamma$ derives from 
\begin{equation}
\gamma = \arccos \frac{\sqrt{1+x^2} -1}{x},
\label{formula2}
\end {equation}
 $x = V/\sqrt{U^2 + Q^2}$,
and $x$ is computed at the left wing of the line.

As an aid in classification of the observed pixels (see Table~\ref{tabellaid}),
we computed the total polarization signal \textit{T} in each image pixel, which is defined as  
\begin{equation}
\mathcal{T} = \int (Q^2 + U^2 + V^2)^{1/2}/I_c d\lambda
\label{formula3}
\end {equation}
where $I_c$ is the intensity in the continuum near the \ion{Ni}{1} line.

Finally, line-of-sight velocities were estimated by Doppler shifts of line cores with respect to the average quiet Sun line center position.

\begin{figure}
\centering{
\includegraphics[width=9cm]{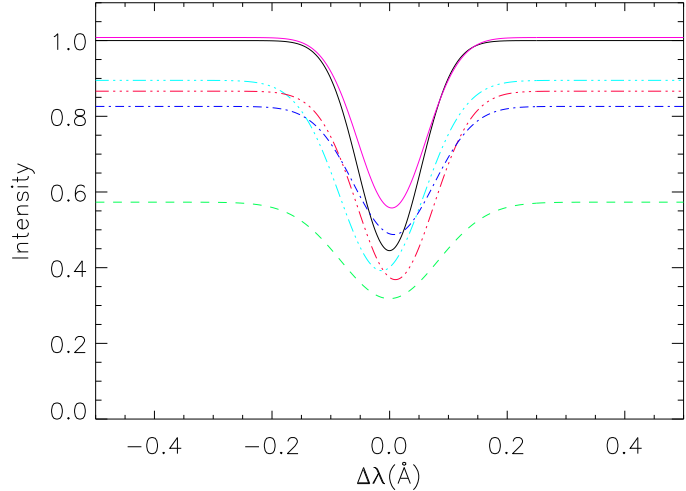}
}
\caption{Line profiles, computed by Gaussian fits to average spectral intensities of various features singled
out on the IBIS observations as indicated in Table 1. Continuous Black: quiet Sun. Continuous Magenta: facula. Dashed Green: umbra. Dot-dashed Blue: pore. Triple-dot dashed Light blue: blue shifted penumbra. 
Triple-dot dashed Red: red shifted penumbra. 
}
\label{fig_prof_obs} 
\end{figure}

Characteristic maps of the continuum intensity $I_C$ of each set 
are shown in Fig.~\ref{fig1}. The FOVs include a large variety of solar features, characterized by different physical properties and therefore different shapes of line profiles. An example of such variety is reported in Fig. \ref{fig_prof_obs}, which shows average line profiles of pixels selected according to the criteria described in Table~\ref{tabellaid}. 

\begin{table}
\begin{center}
\caption{\label{tabellaid} Criteria utilized for the identification of the solar features depicted by masks in Fig. \ref{fig1} and whose line profiles are shown in Fig. \ref{fig_prof_obs}.}

\begin{tabular}{lll}
\hline
\hline
Solar feature & Criterion & Data set \\
\hline
quiet Sun 	  &  $T \leq$  $3.0 m\AA$  			  & \textit{Spot} \\
& & \textit{Faint Plage} \\
facula 		  & $B_{LOS} \geq$ 0.2 kG  			  & \textit{Faint Plage} \\
			  &   and $C_{core} \geq$  10\% 	  &  \\
umbra 		  & $B_{LOS} \geq$  1.5 kG 			  & \textit{Spot} \\
penumbra blue sh. 	  & $ 0.8$ kG $\leq B_{LOS} \leq$ 1.1 kG& \textit{Spot} \\  
			  & and $v_{LOS} \leq  0$ 	 	   &  \\ 
penumbra red sh.   & $ 0.8$ kG $\leq B_{LOS} \leq$ 1.1 kG& \textit{Spot} \\   
    	  & or $v_{LOS} \geq 0$ 		   &  \\ 
pore 		  & $I \leq ave - 5\sigma$ 	 	   & \textit{Pore}\\
\hline
\end{tabular}
\tablecomments{$B_{LOS}$ indicates the magnetic flux strength along the line of sight, 
$C_{core}$ is the intensity contrast in the \ion{Ni}{1} line core that is defined by the ratio $I/I_{QS}$,
where $I_{QS}$ is the quiet Sun intensity,  
$v_{LOS} $ indicate the velocity of plasma along the LOS, 
$ave$ and $\sigma$ are the average 
and standard deviation of all the intensity values measured on the FOV.}
\end{center}
\end{table}

\subsection{Synthetic spectra}
\label{SpecSyn}
\begin{figure} 
\centering{
\includegraphics[width=9cm]{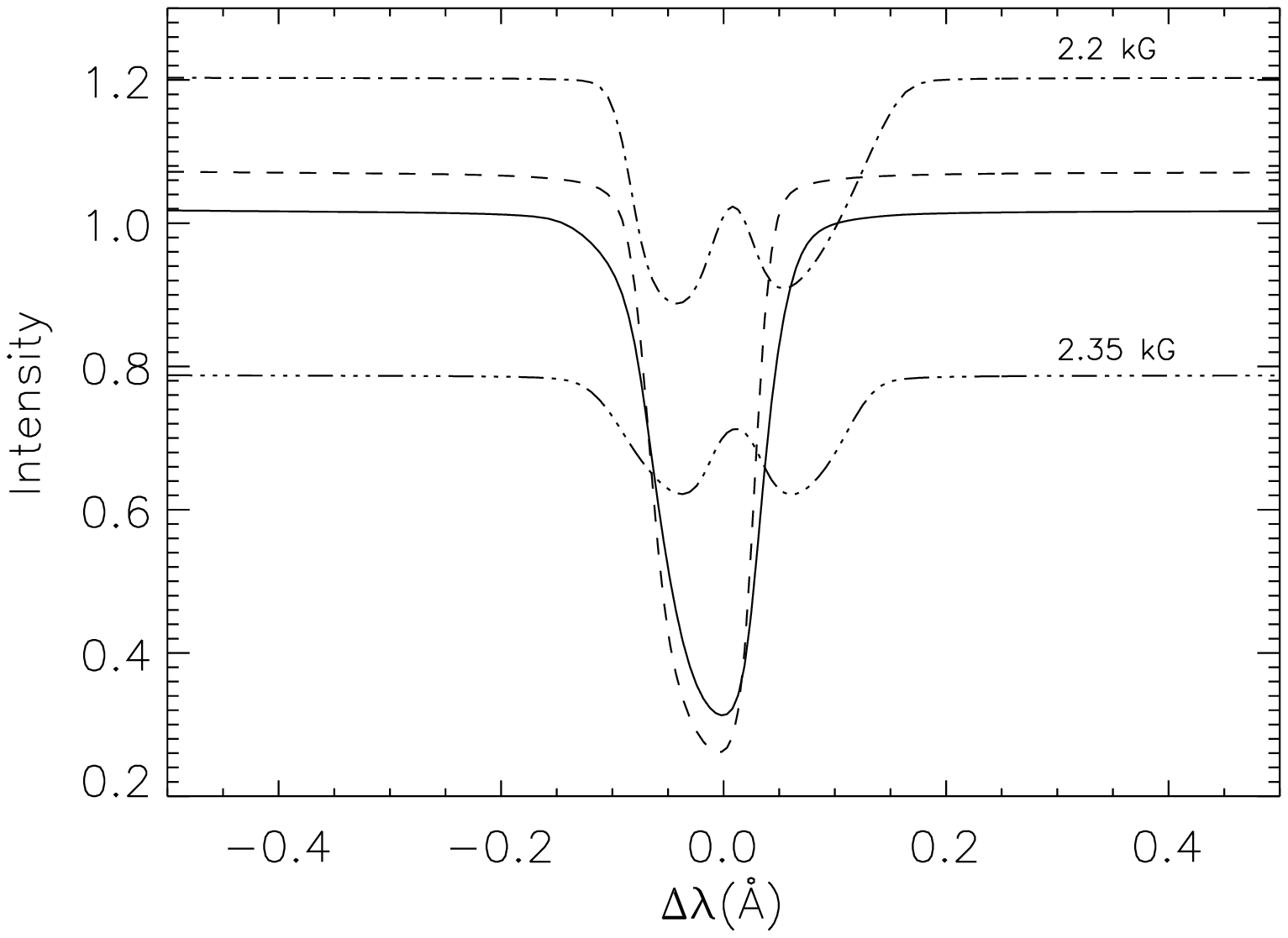}
\includegraphics[width=9cm]{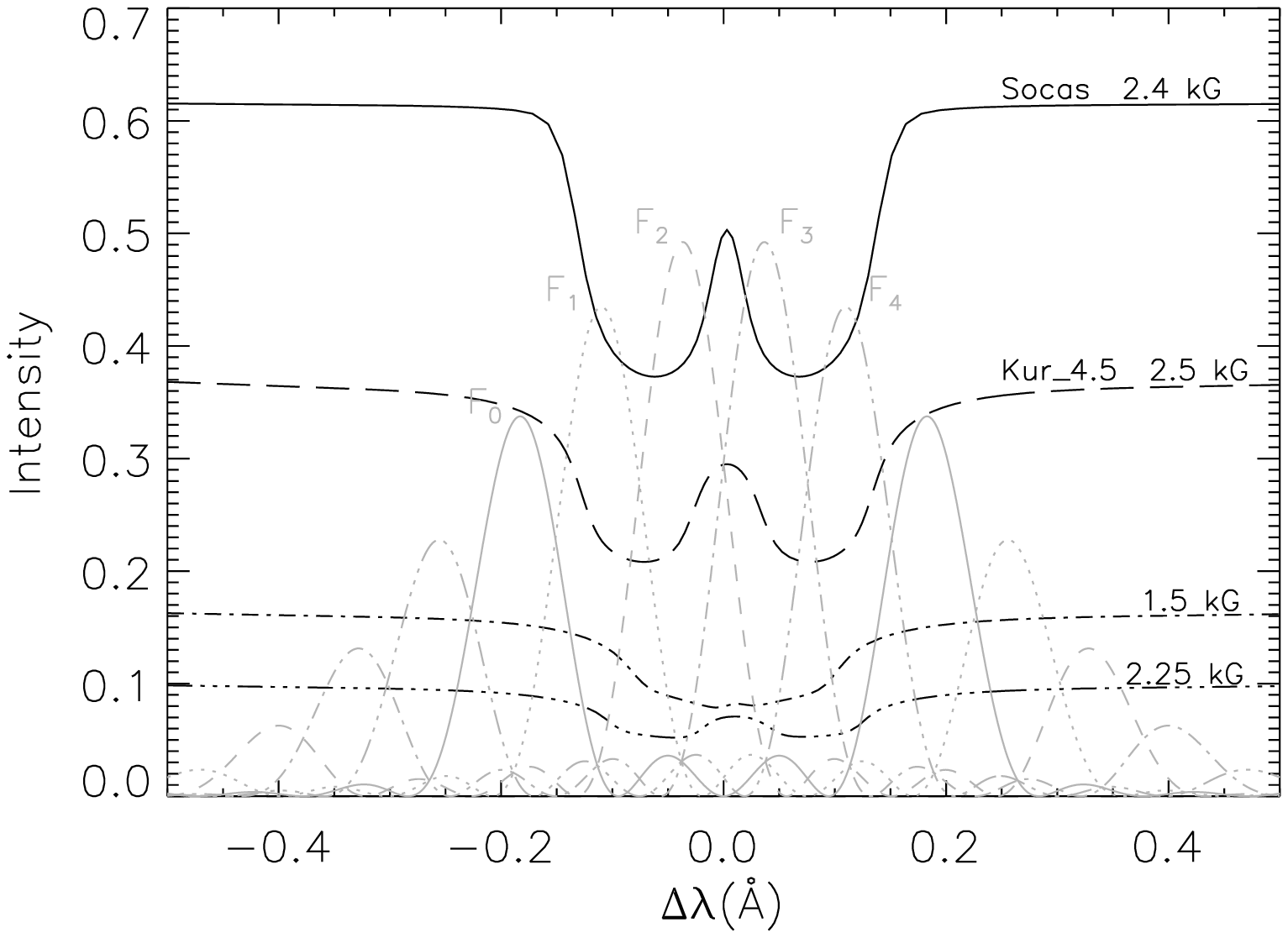}
}
\caption{Line profiles derived from the numerical simulations. Top panel: results from the active region magneto-convection simulation.
Solid: quiet Sun. Dashed: upflow. Dot-dashed: bright magnetic feature (facula). Dot dot dashed: dark magnetic
feature (micropore). 
Bottom panel: results from hydrostatic models and magneto-convection simulations of sunspot. Overlaied profiles show MDI spectral transmission profiles
normalized to fit the figure}.
\label{fig_prof_simula} 
\end{figure} 

To better understand the results obtained from observations and
to distinguish between observational and methodological effects, we
also computed synthetic spectra of the \ion{Ni}{1} line by utilizing
various numerical atmospheres: two two-dimensional cross sections
from a three-dimensional simulation of magneto-convection of an
active region with average vertical field of 250 G
\citep{stein1998}, a vertical cross section of an umbral
magneto-convection simulation \citep{vitas2010}, and various
one-dimensional hydrostatic models.  The one-dimensional static
models we considered ranged from the semi-empirical umbral models E,
L, and M from \citet{maltby1986}, the model by
\citet[][Socas, hereafter]{socas2004} to the most recent radiative equilibrium models of Kurucz
\footnote{http://wwwuser.oat.ts.astro.it/castelli/grids.html}.
In all the Maltby and Kurucz models, we
assumed vertical height-independent magnetic fields. For the Maltby
models, we assumed magnetic field strengths ranging from 2.0 kG to
4.0 kG. The full set of one-dimensional models, which are all
representative of umbral regions, and the corresponding magnetic field strengths are listed in Table
\ref{tablemod}. The field strength of the Socas model listed in this table,
however, is the value at $\tau = 1$ at 5000 \AA, since this model
was used with its original depth-dependent magnetic field
structure. Note that we set the turbulent velocity value in Kurucz
models to zero, since other values in these models are more
representative of non-magnetic and convectively unstable
atmospheres.

\begin{table}
\centering{
\caption{\label{tablemod} One-dimensional atmosphere models considered. Kurucz models are indicated with Kur\_XXX where XXX is the Effective temperature in $10^3$ K units.} 

\begin{tabular}{ll}
\hline
\hline
 Model & B(kG)\\
\hline
Kur\_5.50  & 1.10\\
Kur\_5.25  & 1.70\\
Kur\_5.00  & 2.25\\
Kur\_4.75  & 2.25\\
Kur\_4.50  & 2.50\\
Kur\_4.25  & 3.00\\
Kur\_4.00  & 3.50\\
Kur\_3.75  & 4.00\\
Maltby L  & 2.00\\
Maltby L  & 2.50\\
Maltby L  & 3.00\\
Maltby M  & 2.50\\
Maltby M  & 3.00\\
Maltby M  & 3.50\\
Maltby E  & 3.00\\
Maltby E  & 3.50\\
Maltby E  & 4.00\\
Socas     & 2.3 \\
\hline
\end{tabular}
}
\end{table}

The spectral synthesis was performed in non-local thermal equilibrium (NLTE) with the RH
code by \citet{uitenbroek2002,uitenbroek2003}, and employing the
model atom of \citet{bruls1993}.  The emergent spectra were
calculated at various lines of sight for a 1.\AA\ wide spectral
interval centered on the \ion{Ni}{1} line.  Figure
\ref{fig_prof_simula} shows examples of line profiles derived from the
synthesis in the various atmospheres, all normalized to the
intensity value calculated from the FAL-C model of
\citet{fontenla1993}, which is representative of the quiet
Sun. This normalization provides a common intensity reference value
for the different simulations. Similarly to observations, the line
profiles derived from the synthesis display a large variety of shapes,
because of the different physical conditions in the simulated
atmospheres. The synthetic profiles differ from those deduced from
observations of magnetic features mostly in terms of the line core
reversal. This must be mainly ascribed to the transverse Zeeman
component \citep{stix2004}, which, especially for high values of
magnetic field strength, can considerably modify the shape of the line
core, and the fact that all observations were taken slightly
off disk center.


\section{Simulation of MDI continuum measurements}
\label{MDI_synt_maps}
To test the algorithm for the MDI continuum
proxy we compare the values it predicts with the real continuum values
in the observed and simulated spectra of the \ion{Ni}{1} 6768 line.
We first evaluate the intensities in each of the five
filter positions by multiplying the spectra with the filter passband
and integrating over wavelength, then add the
integrated intensities according to Eq.~\ref{eq:IcontMDI}.
The MDI transmission profiles were centered on the core of the average 
line profile of the Quiet Sun. This was defined as in Table 1 in the case of \textit{Sunspot} and \textit{Faint Plage} dataset, while  for the \textit{Pore} dataset we considered an area in the FOV of 80 square arcsec far from the pores. As for simulations, we centered the transmission profiles to
the core of the line profile derived from the FAL-C model.

Since we generally do not measure absolute intensities in observations,
but rather intensities relative to some average intensity,
we analyze validity of the MDI continuum estimate
by measuring contrast, where contrast is defined as the ratio between the
intensity of an analyzed feature and the intensity of the
quiet Sun. For observed spectra, the contrast was estimated
after compensation for the center-to-limb variation (CLV) of quiet Sun
intensity across the FOV on each frame.  The CLV was derived from the
second order polynomial surface best fitting the intensity pattern of
quiet Sun regions in the image, after masking dark magnetic features
by applying an intensity criterion thresholding, and application of a
smoothing function to pixel intensities. Concerning numerical
simulations, intensity values were normalized by the intensity at the
\ion{Ni}{1} continuum calculated with the FAL-C model for the appropriate disk
positions.
Furthermore, we define as \textit{brightness correction} factor 
to MDI continua
measurements the ratio $C/C_{MDI}$, where $C$ is the contrast value obtained in the continuum
near the \ion{Ni}{1} line (hereafter called the continuum contrast) and $C_{MDI}$ is the contrast derived with
the MDI method (hereafter called the MDI contrast).
In the following we investigate the dependence of this
correction factor on various properties of the analysed regions,
e.g. the contrast, magnetic field strength, and plasma velocity, as
well as observational conditions.

\subsection{Results from observed spectra}
\begin{figure} 
\centering{

\includegraphics[width=4.0cm]{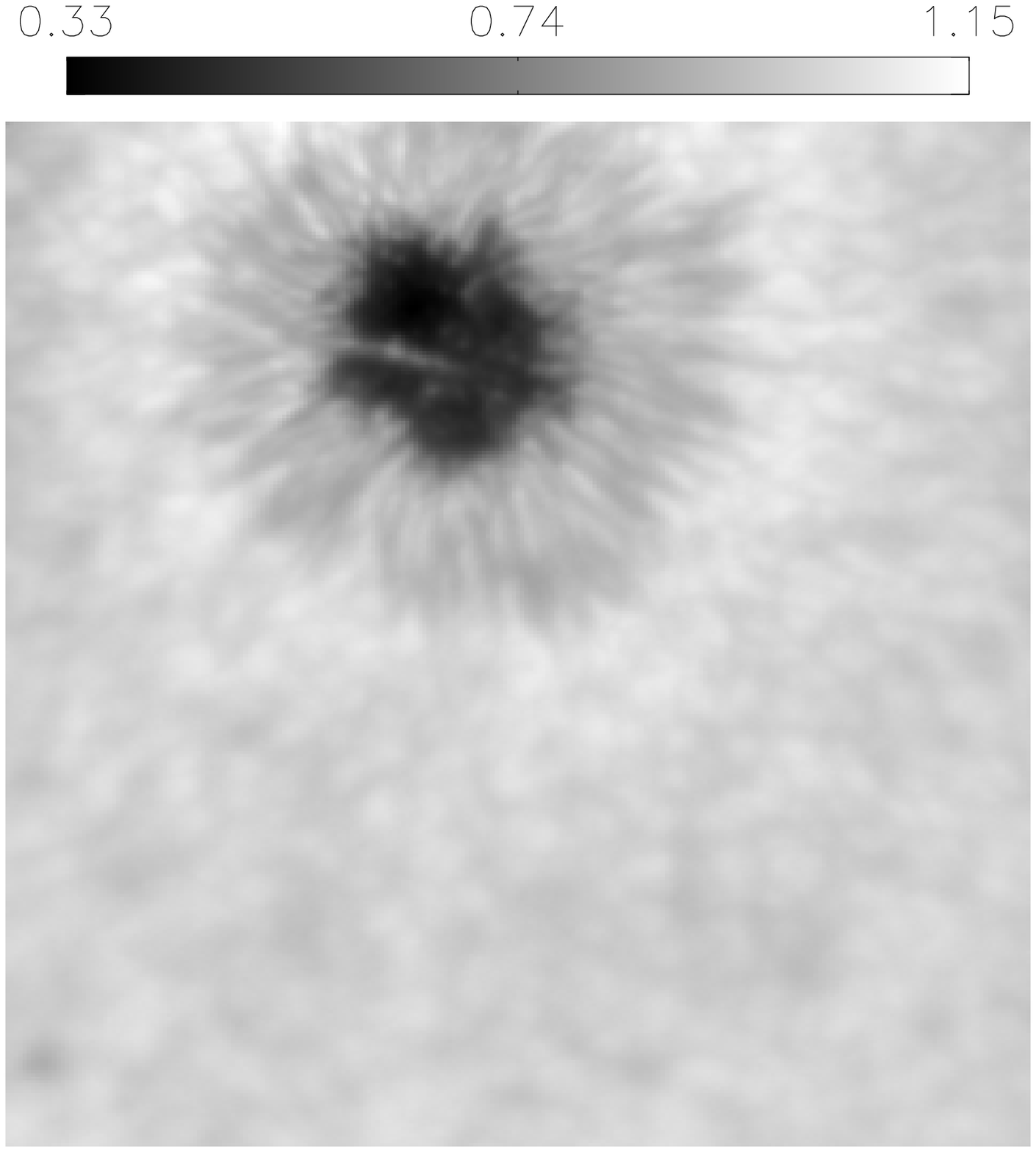}
\includegraphics[width=4.0cm]{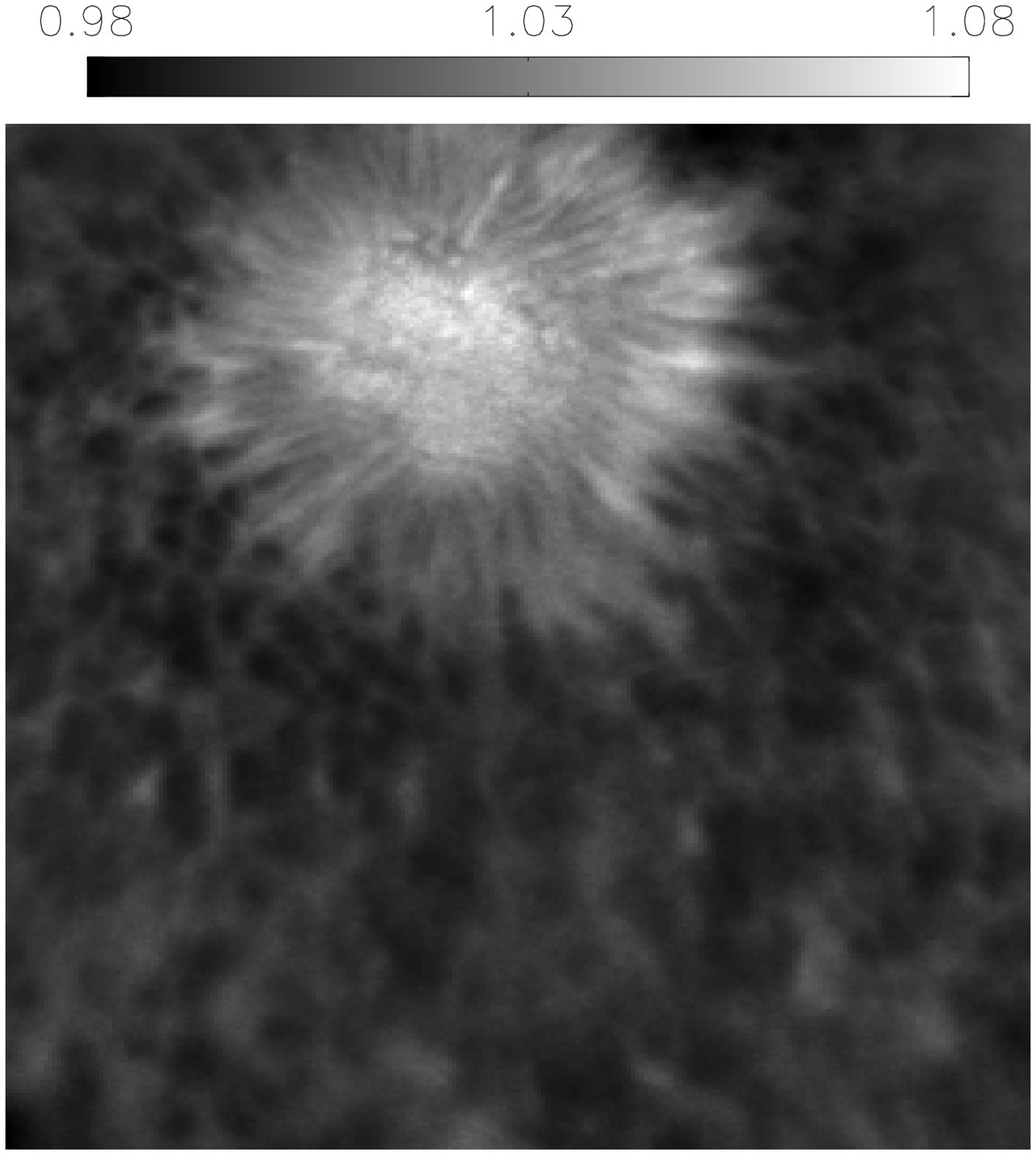}
\includegraphics[width=4.0cm]{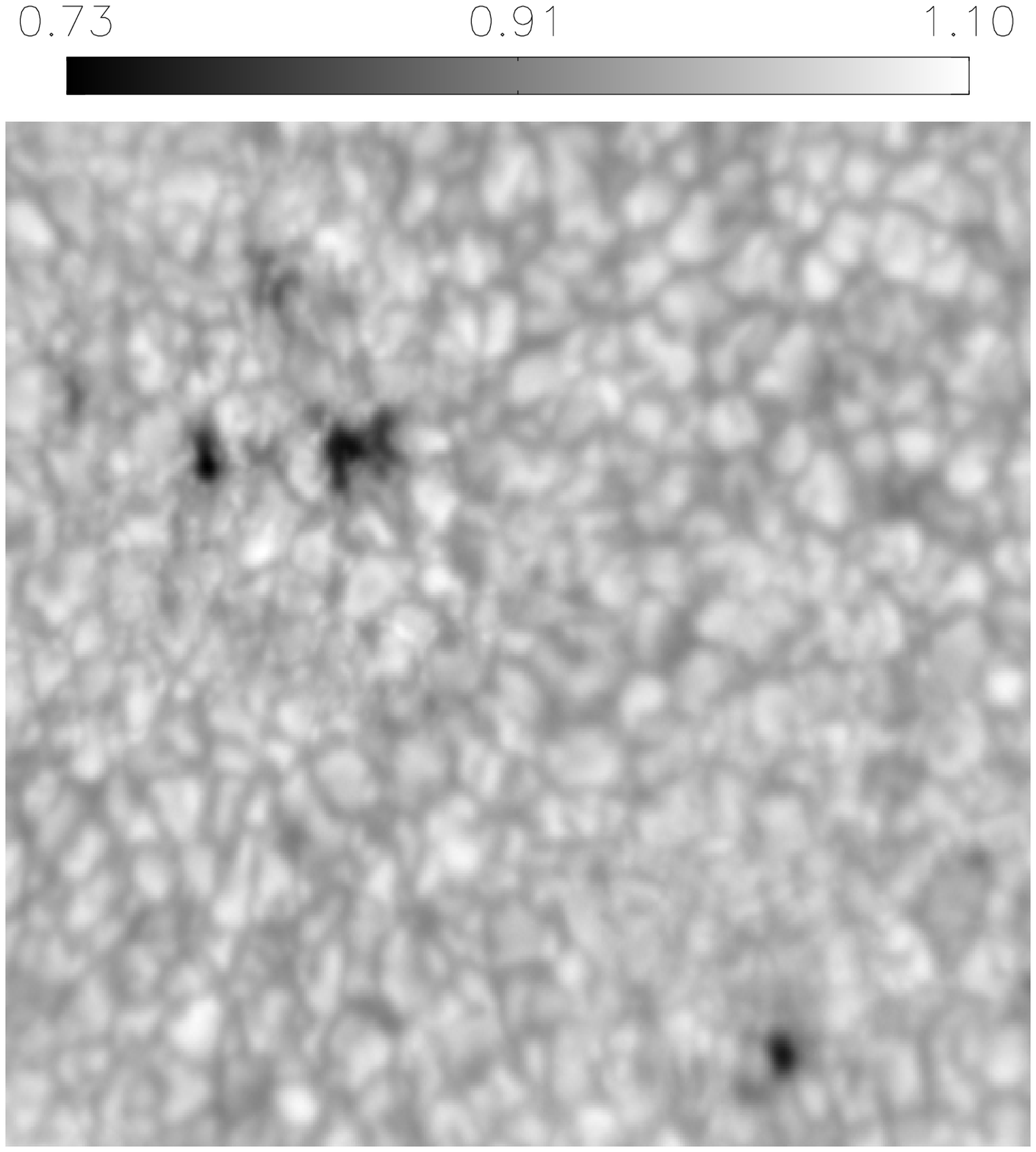}\includegraphics[width=4.0cm]{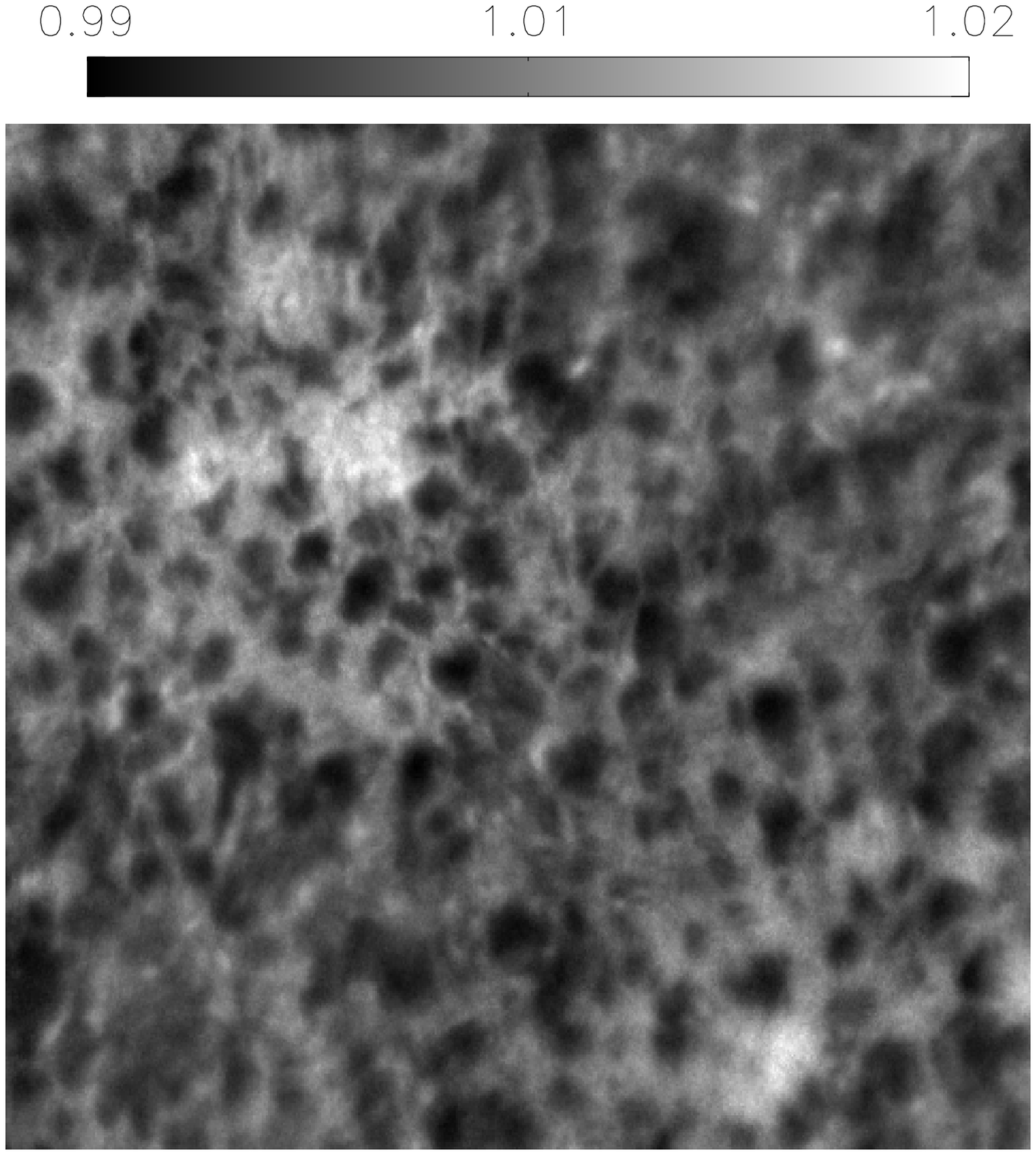}\\

}

\caption{Left panels: continuum contrast maps derived from \textit{Sunspot} (top) and \textit{Pore} (bottom) datasets. 
Right panels: corresponding maps of the brightness correction.}
\label{fig3} 
\end{figure}  

\begin{figure} 
\centering{
\includegraphics[width=9cm]
{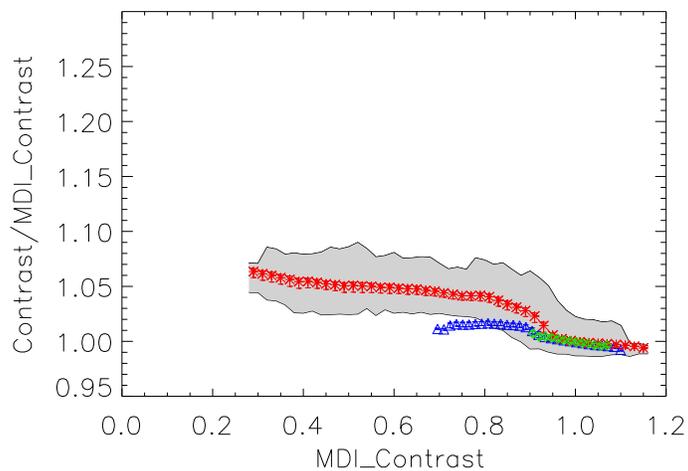} } 
\caption{Dependence of the brightness correction factor on the continuum
  contrast derived with MDI method. Red asterisks, blue triangles, and
  green diamonds show values derived from averages over contrast bins,
  each 0.02 wide, on the \textit{Sunspot}, \textit{Pore}, and \textit{Faint Plage} sets, respectively.  The
  error bars represent the standard deviations of the values in each
  bin. Shaded area delimits the values within  3\% of the maximum of the distribution of values in each bin for the \textit{Sunspot} data set.} 
\label{fig4} 
\end{figure}  

\begin {figure} 
\centering{
\includegraphics[width=9cm]
{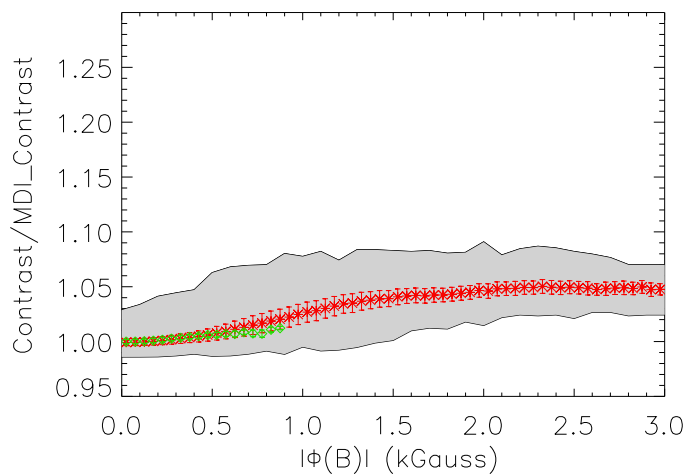}
}
\caption{Dependence of the brightness correction on the magnetic flux
  of the analyzed region. Legend as in Fig.\ref{fig4}.
}
\label{fig_ercontr_mag} 
\end {figure}  

\begin {figure} 
\centering{
\includegraphics[width=9cm]{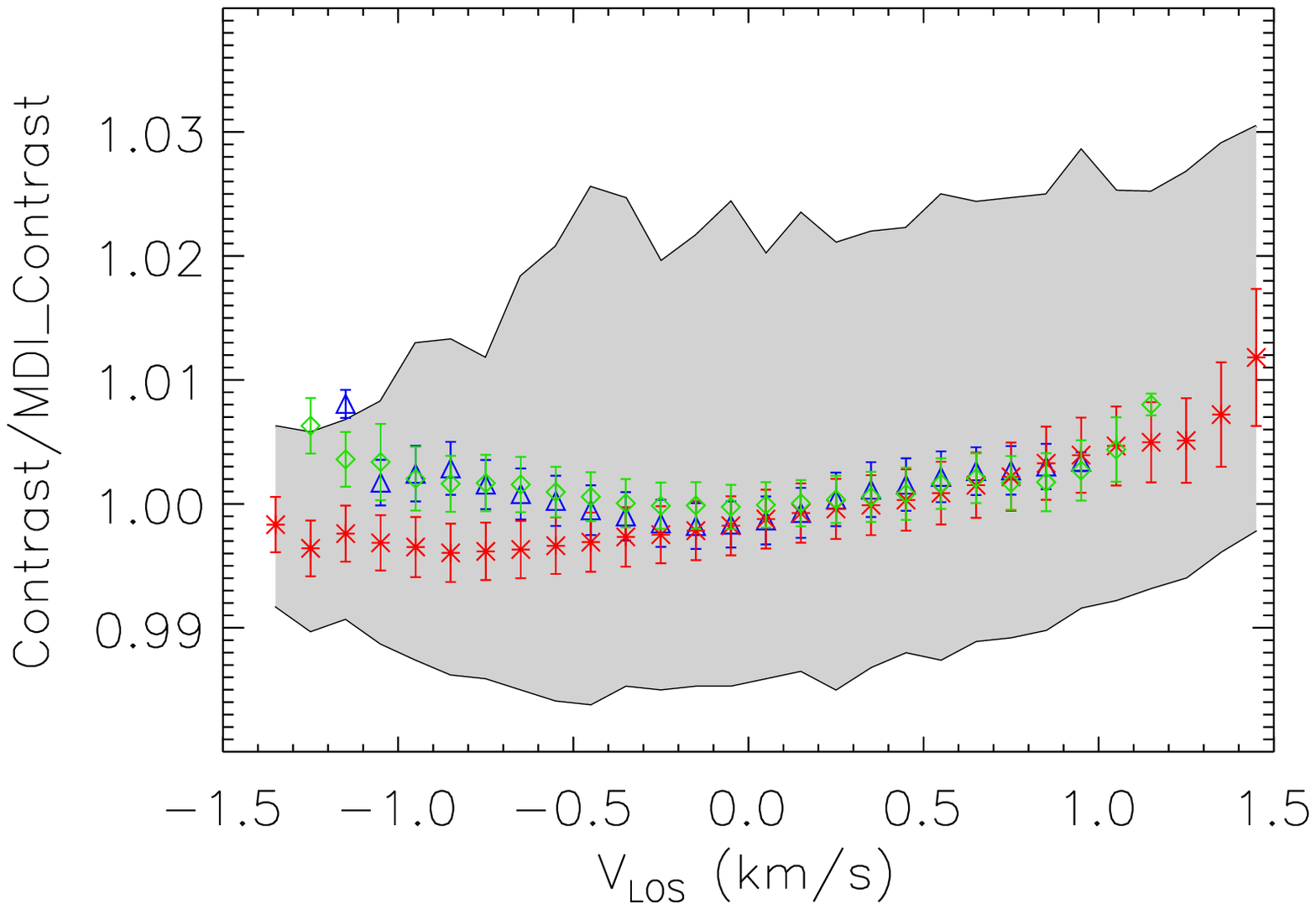}
}
\caption{Dependence of the brightness correction on the LOS velocity for quiet Sun regions identified on the three IBIS dataset. Negative velocities
 correspond to upflow motions, positive to downflow motions. Legend as in Fig. \ref{fig4}}

\label{fig_ercontr_velo} 
\end {figure}  


Figure \ref{fig3} shows maps of the intensity contrast (\textit{left panels}) in the \textit{Sunspot} (\textit{top}) and
\textit{Pore} (\textit{bottom}) sets and the corresponding maps of the
brightness correction to MDI measurements (\textit{right panels}).  We
found that the correction factor is, on average, lower than 2\% of the MDI contrast,
but increases up to 8-10\% in regions with high
magnetic field strength. The figure also shows that the MDI contrast
is, on average, overestimated in upflow regions (granules), while it is
underestimated in regions with downflows (intergranular lanes),
and in particular in regions with strong magnetic fields.
The map of the brightness correction in the \textit{Pore} (\textit{lower right panel})
looks very similar to a map of the inverse granulation as it would
appear in core intensity of a photospheric spectral line, implying
that the correction factor depends strongly on the photospheric
temperature gradient, which is steep over granules and shallow over
inter-granular lanes. The reason that the MDI continuum estimate depends
on this temperature gradient lies in the corresponding widths of the
spectral line profile, which are narrow and deep in granules, and
wide and shallow over intergranular lanes. Profiles that are wider
than that of the average quiet Sun cause an underestimate of the line depth
(Eq.~\ref{eq:linedepth}), and consequentially an underestimate of the MDI
continuum (Eq.~\ref{eq:IcontMDI}). The effect is also visually enhanced by the fact that magnetic field concentrations preferentially reside in downflow regions. This is further explained in 
Section 3.2 below.

Figure \ref{fig4} shows the brightness correction to the MDI contrasts
derived from all the image pixels in all three data sets.
Symbols and error bars of a given colour show the average and standard
deviation of the correction factor derived from a given data set by
considering the MDI contrast values for the set collected into bins
with 0.02 width.  We found that the brightness correction is close to
unity for bright features, because of the opposite contribution from
Doppler shifts and magnetic fields to the line shape, 
though we note
some dispersion of results. On the other hand, the correction factor
increases with the decrease of the MDI contrast derived from the images,
reaching 1.08 for MDI contrasts $\le$ 0.3.

Next we investigated the effects of the magnetic field strength on
MDI contrast estimates in more detail.  Figure \ref{fig_ercontr_mag} shows the
results derived from a sample of \textit{Sunspot} and \textit{Faint Plage} images (7 and 5 IBIS line scans,
respectively).  We found that the brightness
correction increases with the increase of the magnetic field strength,
although a saturation value of $\approx$ 1.04 is reached at magnetic
flux larger than 1.5 kG. In spite of the small standard deviation
values, we also notice a large scatter of results, especially for the
regions with $|\Phi (B)|\leq$ 1.5 kG. In the range 0.7 kG $ \leq |\Phi
(B)|\leq$ 1.2 kG this is due to combined effect of LOS velocity and
magnetic field strength, since this range also comprises pixels belonging
to the sunspot penumbra.  At lower values of magnetic flux the large
scatter of results likely derives from uncertainties in the COG estimations of
magnetic field, which are due to limited spatial resolution and residual cross-talk in Stokes signals. 

Finally, we investigated the effects of LOS velocity on brightness
correction, restricting the analysis to finite ranges of magnetic
field strength values. Results obtained for quiet Sun regions selected
on the three datasets are shown in Fig. \ref{fig_ercontr_velo}.
In agreement with visual inspection of images in Fig. \ref{fig3} the
brightness correction is, on average, larger than unity for downflow regions and
smaller than unity for upflow regions, but with large dispersion. Due to the lower image quality of \textit{Pore} and \textit{Faint Plage} datasets, this effect is more evident on the \textit{Sunspot} data.

\subsection{Results from synthetic spectra}
%
\begin{figure} 
\centering{
\includegraphics[width=9cm]{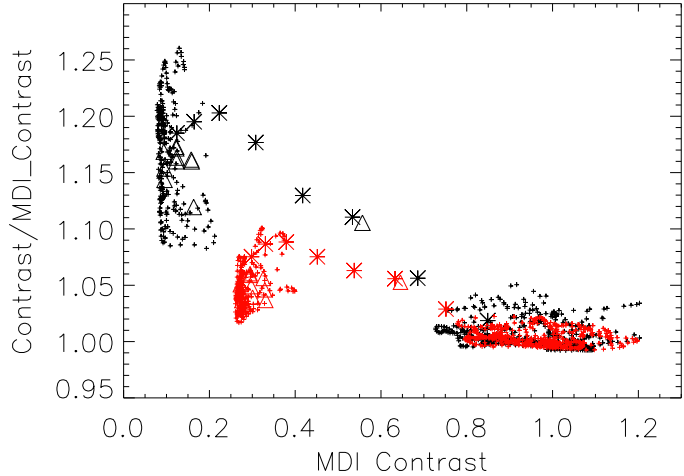}}
\caption{Dependence of the brightness correction to the MDI contrast
  computed for the various atmospheres considered in this
  study. Crosses: results from the magneto-convection calculations. Asterisks: Kurucz
  models. Triangles: Maltby and Socas models. Black: results obtained at disk center. Red: results obtained at disk position
  $\mu$=0.88 after data degradation to account for the finite spectral
  resolution, spurious light and scattered light degradations
  affecting the IBIS data.  }
\label{fig42} 
\end{figure}  

\begin {figure} 
\centering{
\includegraphics[width=9cm]{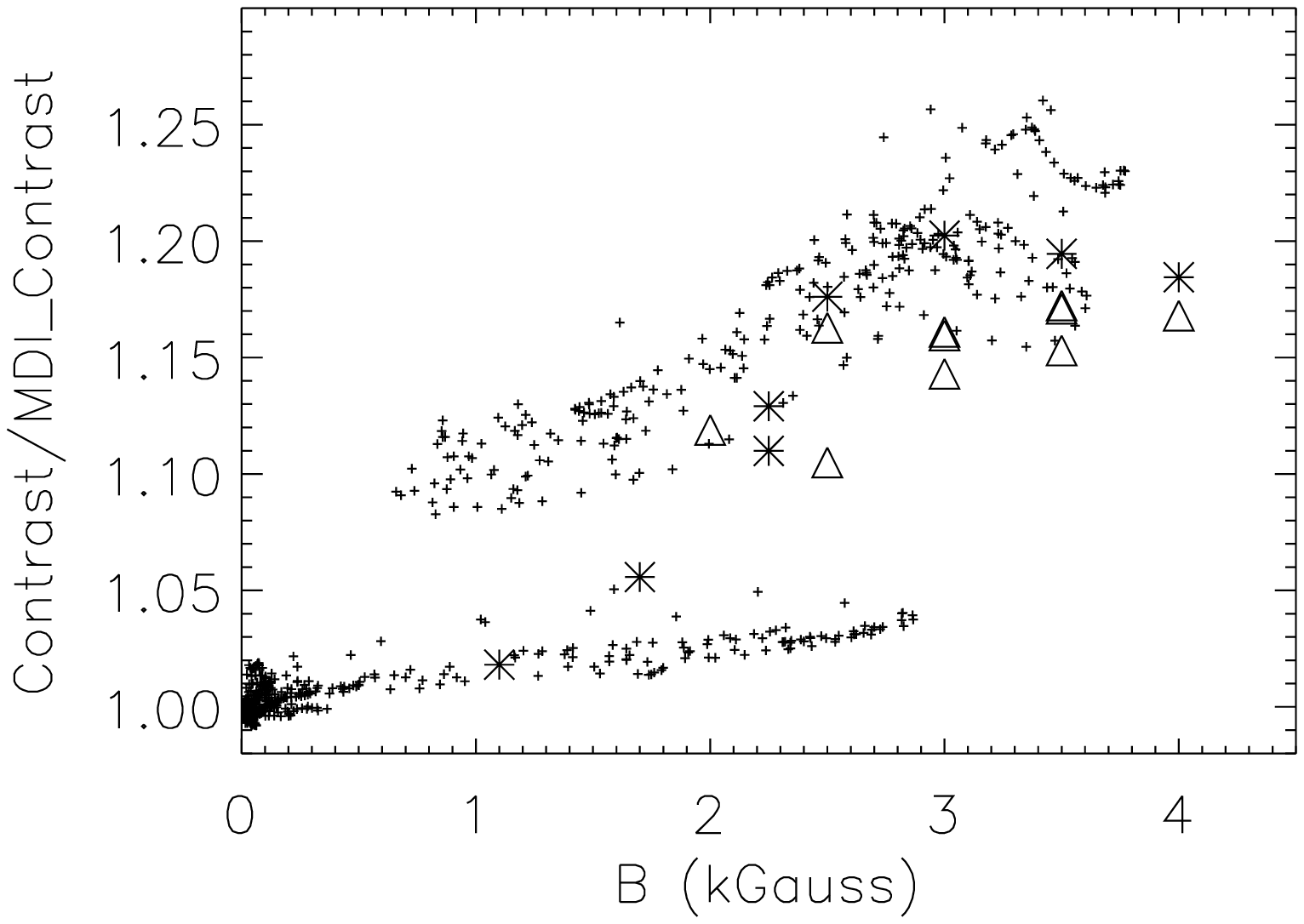}
}
\caption{Dependence of the brightness correction on
the magnetic field strength derived from synthetic spectra. Legend as in Fig. \ref{fig42}}
\label{fig_ercontr_mag2} 
\end {figure}  

Our analysis of observations performed with the observed \ion{Ni}{1}
spectra indicates that
the brightness correction to MDI measurements depends on the solar
feature analysed. This is confirmed by the results derived from
various numerical simulations. Similarly to Fig. \ref{fig4} derived from
observations, Fig. \ref{fig42} shows the brightness correction deduced
from the various atmospheres described in Sect.~2. Here the set of points with contrast close to unity stems from the magneto-convection simulation of active region, the set of points with contrast close to 0.1 stems from the magneto-convection simulation of sunspot umbra and the intermediate contrast points stem from hydro-static models. We found that the
brightness correction is close to unity, with some dispersion of
results due to magnetic field and velocity effects, for MDI contrast
values ranging from 0.8 to 1.2, as also derived from the three IBIS
data sets. However, the correction factor increases up to 1.15-1.25
for MDI contrast $\leq$0.5, i.e., by a larger fraction than derived from observations.
We also found a large dispersion of results from the magneto-convection simulations of
umbral regions, because of the wide range of field strengths in the
analysed atmosphere models. The dispersion of correction values from
simulations is also larger than those from observations.

To assess the importance of observational effects on the results
derived from the three data sets and to place them in context with
the results deduced from model calculations, we computed synthetic
spectra from the various model atmospheres at the heliocentric angles
of our observations (e.g. $\mu$ = 0.88 for the \textit{Sunspot} set) and degraded
the results to account for the finite spectral resolution, spurious
spectral light contamination, and spatially scattered light degradation
affecting the observations.

This latter effect was mimicked by applying
the formula $I_S = (1 -\alpha )I + \alpha I_q$, where $\alpha$
indicates the level of spurious spatial  light contamination, and $I_q$ is the
quiet Sun intensity value. Estimation and compensation on high spatial
resolution data for scattered light results uncomplete, even employing the most sophisticated restoration algorithm available \citep[e.g. MOMFBD][]{scharmer2010}. In the lack of measurements, we
assumed the value $\alpha = 0.2$, which is within the range
of values (0.17-0.25) found by \citet[][]{briand2006} for IPM/THEMIS
data. The results obtained indicate that the observational conditions
and degradations above listed all affect the brightness correction,
specifically by decreasing its value. The difference between the three IBIS datasets must therefore be ascribed to the
different image qualities.  We also note that, among the observational
effects considered, the brightness correction is affected the most by
spurious spatial light contamination, which also contributes to a smaller
dispersion of the results from observations \citep[][]{mathew2007,criscuoli2008}.

Figure~\ref{fig_ercontr_mag2} shows the dependence of the brightness
correction on the magnetic field strength in the synthetic
atmospheres.  The results deduced with simulations are qualitatively
in agreement with those derived from observations, though the
saturation value is reached only at B = 3 kG and the dispersion of results
is larger than obtained from observations as a result of image
degradation by limited spatial resolution and spatially spurious light.
The plot clearly shows a double distribution:
one for which the brightness correction value is smaller than 1.05 and
one for which it is higher. We found that data from the active region magneto-convection simulation belong to the first group, whereas the second group is
populated by sunspot models (with the exception of the
two hottest Kurucz models) and the umbral simulation.

We found that the MDI continuum estimate method depends in a complicated
way on both temperature and magnetic field intensity. The larger
the magnetic field strength, the wider the line profile.
If the line profile becomes so wide and/or shifted that
it reaches into (one or two of) the outermost transmission $F_0$ passbands, 
the continuum (Eq.~\ref{eq:IcontMDI}) is underestimated, resulting in a brightness
correction larger than one.
This effect is also discussed in \citet{mathew2007} and
illustrated in their Fig. 7. These authors also notice that for the
largest magnetic field values considered in their study, a saturation
of the error occurs because of the filling in of the line core,
as is also clear in the line profiles in Fig.~\ref{fig_prof_simula}.
By contrast, an increase in the effective temperature
narrows the line and increases its depth \citep[][ their figure 8]{mathew2007}.
As a consequence, for a given value of magnetic field strength the brightness
correction is smaller for faculae rather than for sunspots. On the other hand,
at the saturation limit, where the field mostly affects
the width of the line, the error in the MDI continuum estimate increases
with the effective temperature as the line narrows. 
We also note that the brightness correction factor for the synthesized spectra
(Fig.~\ref{fig_ercontr_mag2}) seems less affected by saturation
than the one for the observed spectra (Fig.~\ref{fig_ercontr_mag}).

\begin{figure} 
\centering{
\includegraphics[width=9cm]{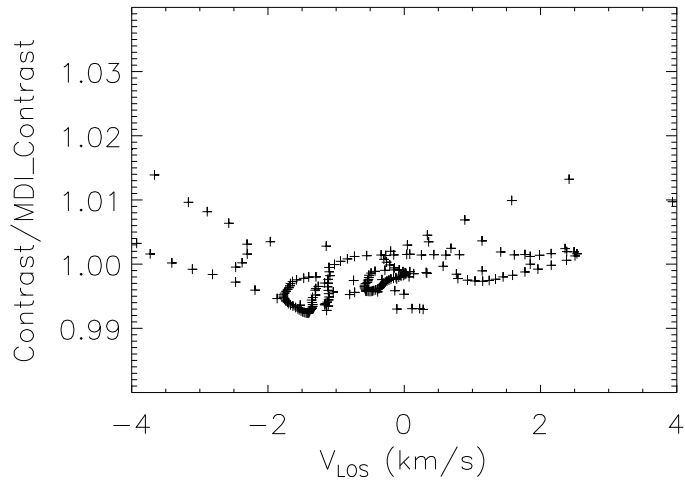}}
\caption{
 Dependence of the brightness correction to MDI measurements on the
 LOS velocity for quiet regions from magneto-convection simulation of active region. }

\label{fig_veloquiet} 
\end{figure} 

Next, we investigated the dependence of the brightness correction on
the LOS velocity of the plasma. We found that, on average, the brightness
correction is larger than unity and increases with the magnitude of the LOS
velocity. This is illustrated in Fig.\ref{fig_veloquiet}, which shows
the results derived for quiet-Sun portions (defined here by those
pixels with $B\leq$ 50 G) of the magneto-convection simulations representative of
active region. As we can see, data from simulations allow to investigate a wider range of velocities with respect to observations \citep{cauzzi2006}. When restricting the analyses to the range from -2 km/s to 2 km/s we note the same asymmetry we found for observations, namely  a brightness correction smaller than unity for upflows and larger than unity for downflows. 
This asymmetry is the result of the correlation
between temperature gradient and up- or downflows in the granulation.
Granules correspond to upflows (negative $v_{\mathrm{LOS}}$) and 
have steep temperature gradients resulting in narrow and deep profiles.
The shallow temperature gradient over inter-granular lanes, where
most downflows occur, leads to broader profiles.
The spacing of the MDI filter positions is chosen such that it
matches the width of the average quiet-Sun profile of the Nickel line
in the sense that Eq.~\ref{eq:linedepth} gives the proper line depth.
For substantially narrower profiles the filter positions $F_1$ and $F_4$
sample close to continuum values increasing the line depth estimate,
overestimating the continuum. This leads to the correction factors
below unity for small upflows.
For larger line shifts to either sides, but already for smaller velocities
in the downflows with their wider profiles, the correction
factor is larger than one, because $F_1$ and $F_2$ (for blue shifts),
or $F_3$ and $F_4$ (for red shifts) both end up close to the core,
minimising their difference in sampled intensity, underestimating
the line depth, and underestimating the continuum level.

\begin{figure} 
\centering{
\includegraphics[width=9cm]{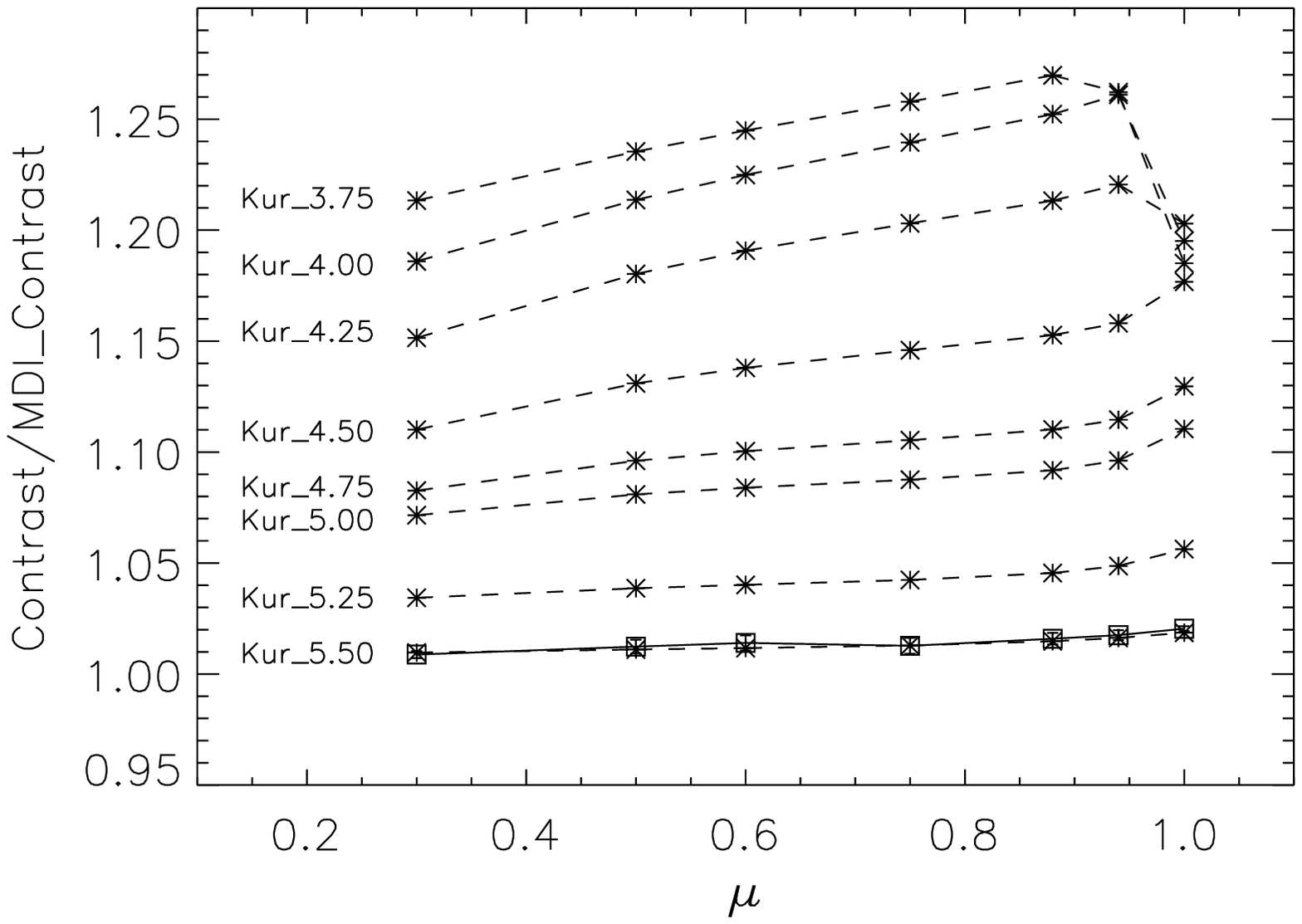}}
\caption{
Center to Limb variation of the brightness correction for various magnetic features investigated through synthetic spectra.
Squares: faculae, from magneto-convection simulation. Asterisks: sunspots, from Kurucz models.}   

\label{fig_CLV} 
\end{figure} 

Finally, we considered the dependence of the brightness correction on
the position of the analysed feature on the solar disk. To this
aim, we computed the synthetic spectra for 13 different viewing angles
from various model atmospheres. In particular, we employed all the
static sunspot models as well as the magneto-convection simulations representative of an active region.
The Nickel line forms low enough in these models so that the vertical
domain includes its formation height, even at high inclinations of the LOS. 
Figure \ref{fig_CLV} shows the correction factor for the different
Kurucz models with cooler models and stronger fields (both leading
to wider profiles, and larger correction factors) towards the top.
For the models with $B \le$ 3 kG (or $T_{eff} \geq 4500$ K), the brightness
correction monotonically decreases with the decrease of the
heliocentric angle, because the width of the line decreases
with the line-of-sight component of the magnetic field, which
changes linearly with $\mu$ for the assumed vertical fields.
For models with $B >$ 3 kG ( $T_{eff} \leq 4250$ K), the CLV of the correction factor 
increases strongly with decreasing $\mu$ from 1.0 to 0.9,
and then decreases again monotonically with decreasing $\mu$.
The initial rise is the result of the transverse magnetic
Zeeman effect, which produces intensity decrease of the core of the profile for
positions close to disk center, thus creating characteristic double-peaked profiles. These quickly disappear 
for $\mu < 0.9$.
Figure \ref{fig_CLV} also shows the CLV of the correction
factor for faculae (\textit{squares}), designated as those areas that have a field
strength larger than 0.2 kG and contrast greater than one)
in the magneto-convection active region simulation. These facular regions show
very little variation in their required correction factors accross
the disk.

\section{Sensitivity to passband uncertainties}
\begin{figure} 
\centering{
\includegraphics[width=9cm]{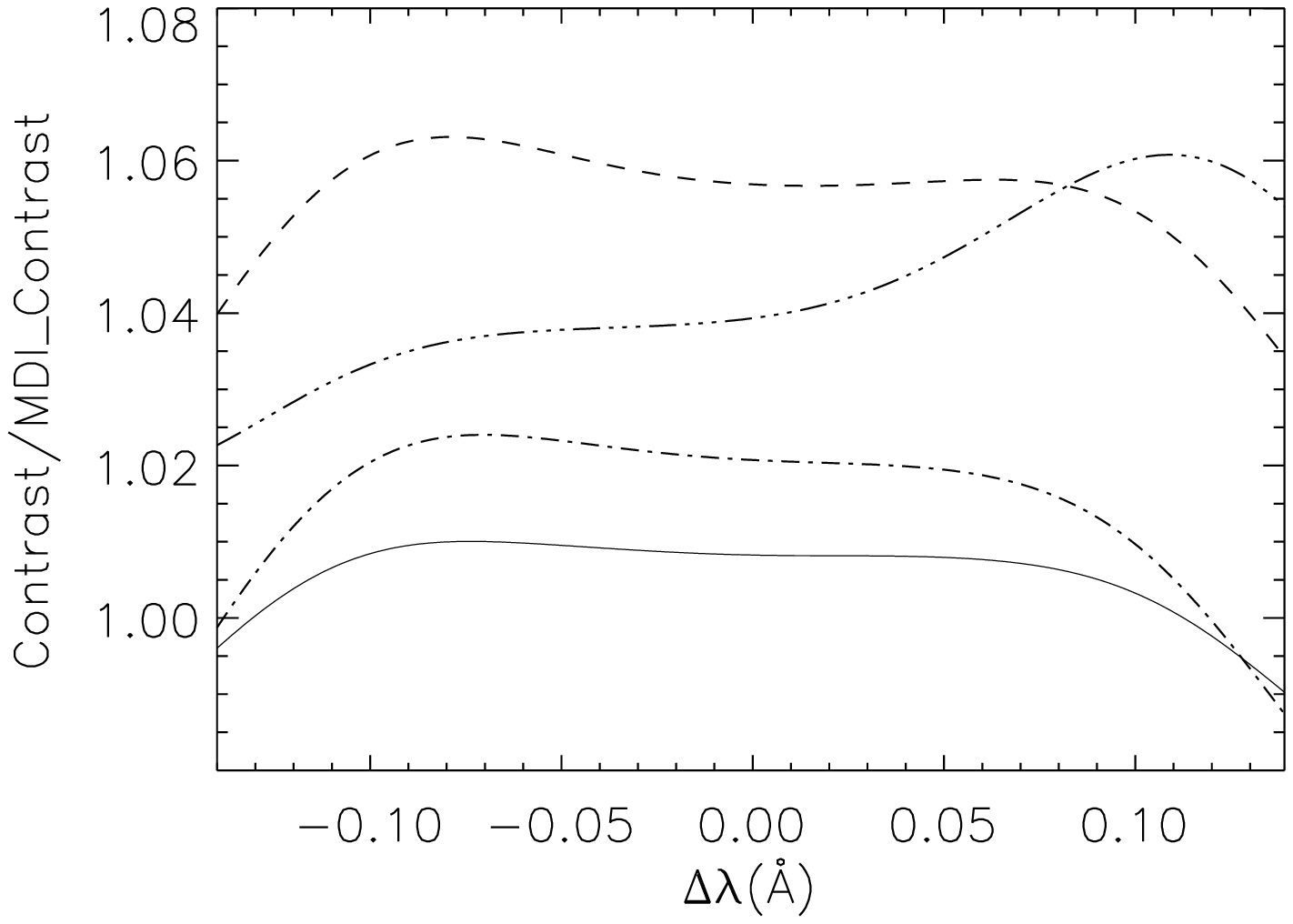}
\includegraphics[width=9cm]{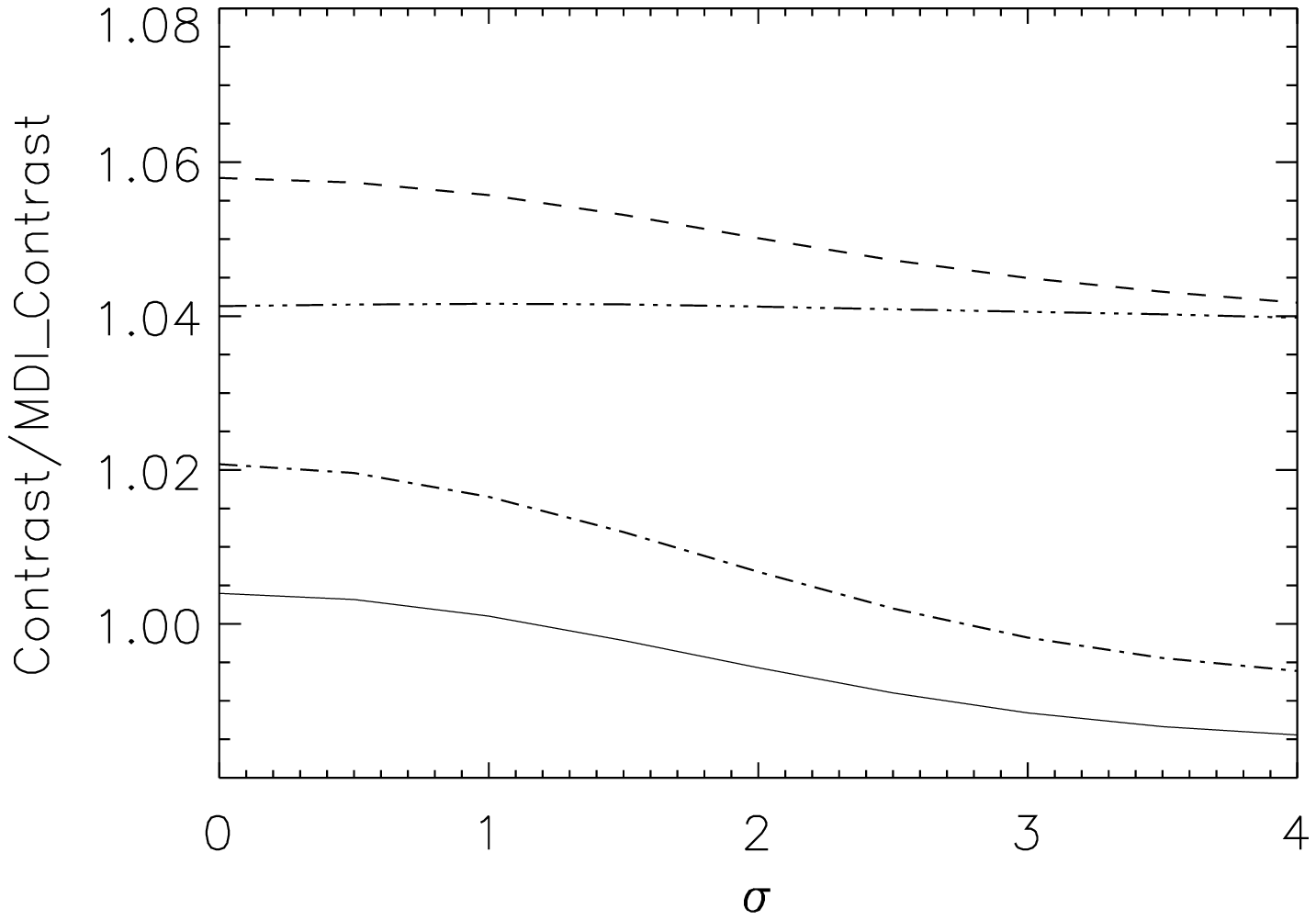}}
\caption{
Variation of brightness correction of features identified on IBIS data (see Table \ref{tabellaid})
with the filter drifts (top panel) and modifications of shapes of
transmission profiles (bottom panel) considered in this
study. Continuous line: plage region. Dashed line: umbra. Dot-dashed line: pore. Triple-dot-dashed line: penumbra.
}
\label{fig_aber_shift} 
\end{figure} 

To assess the importance of an inaccurate knowledge of the filter
characteristics, we analysed the dependence of the brightness correction
derived from observed spectra on potential inaccuracies in the
precise tuning and width of the MDI passbands.
In particular, we computed the variation of the brightness correction for
displacements of the $F_0 \ldots F_4$ passband pattern ranging from -0.14 \AA~ to +0.14 \AA~ from its
nominal position.
These values correspond to Doppler shifts of $\pm\ 6$ km/s 
and are 1.5 times larger than the ones considered in the calibration of
the MDI measurements \citep[][their Fig. 12]{scherrer1995}.
We also investigated the sensitivity of the required brightness
correction to variations of the width of the MDI passband respect to the nominal value of 94 m\AA.

Figure \ref{fig_aber_shift} displays the resulting brightness correction
as a function of displacement of the MDI pass band pattern, when
considering the same average observed line profiles for quiet Sun, plage,
pore, umbra, and the most blue-shifted penumbra (see Fig.~\ref{fig_prof_obs}).
We found that, for moderate displacement values, the brightness correction decreases
for displacements in the same direction as the line profile Doppler shift,
while it increases for shifts in the opposite direction.  This is
particularly clear for the penumbral profile (\textit{triple-dot-dashed curve}).
For large displacements all the curves in Fig. \ref{fig_aber_shift} show a
trend towards unity, because the offset between the line center and 
transmission pattern becomes so large that the passbands sample more
and more true continuum. Note also the oscillatory variations of the curves, that, as pointed out by \citet{wachter2008}, have a period proportional to the spacing of the transmission profiles of the filters.  

{Next, we computed variation of the brightness correction for different
assumed widths of the tunable MDI passband by convolving the nominal
profiles with Gaussians of widths that are multiples of the width of the profiles.
This last quantity was estimated fitting the highest lobe of $F_0$ with a Gaussian function. 
Figure \ref{fig_aber_shift} shows the results for the various
solar features.
We found that the brightness correction decreases with the increase of
the filter profile widths, and eventually drops below
unity for the pore and facular profiles.
This is the result of the transmission profile becoming wider than
the line profile (see also the discussion in 3.2).
Note also that the brightness correction for the penumbra is the least
affected by the widening of transmission line profiles, since its value
is mostly determined by the shift of the spectral line.



\section{Discussion and Conclusions}

We have presented a study of the accuracy of the MDI continuum intensity
measurement, with the help of high spatial resolution
spectro-polarimetric observations acquired with the IBIS/DST and
outcomes from numerical simulations representative of different solar
regions. In order to provide the reader with a quantity that clearly
indicates the MDI uncertainties and allows their compensation on
measurements, we defined as brightness correction the ratio between
the contrast value in the continuum and the one estimated through MDI
method and investigated the variation of this quantity with the physical
properties of analyzed solar regions.

We found that the MDI brightness correction, on average, increases
with the increase of the LOS magnetic field strength, the decrease of
temperature, and the increase of LOS velocity.  To summarize, the
contrast is overestimated in quiet upflow
regions, and is underestimated in magnetic and quiet downflow
regions. The amount of correction is less than 1\% in quiet regions and increases with the magnetic field
strengths, reaching values up to 25\% of the MDI continuum value in sunspots.
However, line saturation effects limit the MDI error in
regions with highest magnetic field strengths. 

Results derived from observations agree with those deduced
from simulations, when accounting for the observational conditions
and instrumental degradations affecting the observations.
Among the effects considered, spatially spurious light is the one that affects
the most the brightness correction values. In the lack of a reliable estimate on our observational data,
we therefore refrained from a direct comparison of results from MDI and IBIS
data. On the other hand, we did not find any significant variation of
the brightness correction values when degrading the results from IBIS
data to the spatial resolution of MDI, thus suggesting that brightness
correction is not significantly dependent on the spatial resolution of
the observations.

Note that the brightness correction values that we found correspond to errors  much higher than  the 0.2\% quoted in the instrument description of
\citet[][p.~163]{scherrer1995}.

The brightness correction factor derived from synthetic spectra
also depends on the position on the solar disk in different manners
for the different features analyzed.  We found that the brightness
correction for sunspot models with strong field ($B_{\mathrm{LOS}} \ge 3$ kG)
varies by up 2\% for disk positions corresponding to $0.9 \leq \mu \leq 1$, reaching
a maximum value for $\mu = 0.9$, and then decreasing towards the
limb. For weaker fields the variation across the disk is much less
(Fig.~\ref{fig_CLV}).
Also for quiet Sun, faculae and micropores the variation in
brightness correction is small, below 1\%, across the disk.

Finally, we investigated effects of possible filter degradations on
MDI continuum contrast estimates. We estimated variations
with the shift of the passband pattern and with the change
in the width of the transmission profiles of the instrument.
Results
are summarized in Fig. \ref{fig_aber_shift} for various features
observed on the solar photosphere. We found that for small amounts
([-0.03,0.03] \AA) of transmission profiles displacement the
variation of the brightness correction is negligible for features not
affected by Doppler shifts (i.e. faculae and quiet sun). For line
profiles shifted with respect to quiet sun line center, a shift of
filter center in the same direction as the line Doppler shift
causes a reduction of the amount of correction, whereas a displacement
in the opposite direction causes an increase.
In addition, we found that broadening of the passband causes a
decrease of the amount of correction when the width of the line
profiles is larger than that of the passband. For transmission line
profiles wider than the profile of the solar feature under study,
the contrast is overestimated and the amount of deviation of the
correction coefficient from unity increases with the increase of the
filter degradation.

Our results are qualitatively in agreement with those presented by
\citet{mathew2007}, who applied MDI procedure to synthetic line
profiles obtained from \citet{kurucz91} atmosphere models. Nevertheless, a comparison of results shows that they
obtained brightness correction larger than the ones we have
presented. These discrepancies must be ascribed to the differences in model atmospheres employed, assumptions for atomic parameters and micro and macro velocity values. Due to the arbitrariness of the choice of these quantities inherent to 1D models, this analyses should be repeated by a direct comparison of MDI data with observations in the \ion{Ni}{1} continuum properly compensated for scattered light, or with the employment of magneto-convection simulations encompassing a larger variety of sunspots.   

Results presented in this study show that photometric measurements
obtained with MDI data should be revised by taking into account the
brightness correction factors presented. Due to the dependence of this
factor on observational conditions, continuum images should be
compensated for spatial scattered light before any correction is
applied. Since MDI Point Spread Function varies with time, this result
contributes to question studies based on the temporal variation of
photometric properties of solar features derived from MDI data
analyses. 

Finally, our study proofs forward modeling as a powerful tool also for instrumental calibration. In particular, results presented here are of interest for the interpretation of data acquired with the Helioseismic and Magnetic Imager onboard the Solar Dynamic Observatory.

\acknowledgments
 We thank Richard Wachter and Kevin Reardon for providing us with one of the IBIS datasets. This study was
supported by the Istituto Nazionale di Astrofisica (PRIN-INAF-07), the
Ministero degli Affari Esteri (Bando 2007 Rapporti bilaterali
Italia-USA), and the Agenzia Spaziale Italiana
(ASI/ESS/I/915/01510710). The NSO is operated by the Association of Universities for Research in
Astronomy, Inc. (AURA), for the National Science Foundation. IBIS has been built by INAF/Osservatorio Astrofisico di Arcetri with
contributions from the Universities of Firenze and Roma ''Tor Vergata'',
the National Solar Observatory, and the Italian Ministries
of Research (MUR) and Foreign Affairs (MAE).


\clearpage

\end{document}